\documentclass[runningheads,a4paper]{llncs}

\usepackage{amssymb}
\usepackage{graphicx}
\usepackage{verbatim}
\usepackage{xcolor}
\usepackage[hidelinks]{hyperref}
\hypersetup{
  colorlinks,
  citecolor=blue,
  linkcolor=blue,
  urlcolor=blue}
\usepackage{url}

\urldef{\mailsa}\path|ana.tomeklock@tuni.fi, bssantana@inf.ufrgs.br, juho.hamari@tuni.fi|

\newcommand{\keywords}[1]{\par\addvspace\baselineskip
\noindent\keywordname\enspace\ignorespaces#1}

\begin{document}

\mainmatter 

\title{Ethical challenges in gamified education research and development: An umbrella review and potential directions}
\titlerunning{Ethical challenges in gamified education research and development}
\toctitle{Ethical challenges in gamified education research and development}

\author{Ana Carolina Tomé Klock$^1$ \and Brenda Salenave Santana$^2$ \and Juho Hamari$^1$}
\authorrunning{Klock, Santana and Hamari (2022)}
\tocauthor{Klock, Santana and Hamari (2022)}

\institute{$^1$Gamification Group, Tampere University, Finland\\
$^2$Federal University of Rio Grande do Sul, Brazil\\
\mailsa}

\maketitle

\begin{abstract}
Gamification is a technological, economic, cultural, and societal development toward promoting a more game-like reality. As this emergent phenomenon has been gradually consolidated into our daily lives, especially in educational settings, many scholars and practitioners face a major challenge ahead: how to understand and mitigate the unethical impacts of gamification when researching and developing such educational technologies? Thus, this study explores ethical challenges in gamified educational applications and proposes potential solutions to address them based on an umbrella review. After analysing secondary studies, this study details and proposes recommendations on addressing some ethical challenges in gamified education, such as power dynamics and paternalism, lack of voluntarity and confidentiality, cognitive manipulation, and social comparison. Research and development decision-making processes affected by such challenges are also elaborated, and potential actions to mitigate their effects in gamification planning, conducting and communication are further introduced. Thus, this chapter provides an understanding of ethical challenges posed by the literature in gamified education and a set of guidelines for future research and development.

\keywords{Gamification, Teaching-learning processes, Virtue ethics}

\end{abstract}

\section{Introduction}

Using game design elements to promote game-like experiences throughout many daily tasks -- namely gamification -- has garnered growing attention among scholars and practitioners over recent years \cite{Hamari2014}. Gamification is an emergent phenomenon in multiple domains, having a meaningful role in upholding many Sustainable Development Goals (SDG), such as good health and well-being (SDG3) \cite{Johnson2016}, decent work and economic growth (SDG8) \cite{Warmelink2020}, and climate action (SDG13) \cite{Galeote2021}. Quality education (SDG4) is prominent for gamification among this variety of possibilities \cite{Koivisto2019}, since it offers a gameful way to engage and inspire students during the teaching-learning process. Consequently, there is a continuously increasing interest in both uses and implications of gamified education \cite{deSousa2014}. 

Nevertheless, despite contributing towards several improvements to the educational field, gamification also introduces adverse effects when not suitably applied \cite{Toda2017}. For instance, while it aims to promote more appealing and rewarding learning, gamification affects diverse students in distinct ways \cite{Smiderle2020}. Such individual differences raise questions on how scholars and practitioners should consider and handle data regarding personal characteristics and preferences -- questions that are ethical by nature. Thus, understanding and promoting ways to ethically research and develop gamified applications in all fields, but especially in education, is an essential question yet to be addressed.

Towards an answer to such a question, this chapter investigates and discusses ethical challenges of gamification research methodology and application development in teaching-learning processes. This study is organised as follows: Section \ref{sec:background} provides an overview of normative ethics and its philosophical theories, as well as their relation to gamification research and development, and further decision-making processes regarding planning, conducting and communicating gamification outcomes. Section \ref{sec:methology} describes the methodology, research questions, inclusion criteria, search process, screening procedure and data extraction plan. Section \ref{sec:results} details bibliometric information of the secondary studies, while elaborating on how to make ethical gamification and how to make gamification ethical. Finally, Section \ref{sec:conclusion} concludes this chapter by presenting the final remarks and limitations.

\section{Background}\label{sec:background}

Ethics is an extensive philosophy branch that analyses and conceptualises moral behaviours and determines right from wrong through multiple perspectives (e.g., meta-ethics, normative ethics, and applied ethics) \cite{Smith2021}. From these sub-branches, this work focuses on the normative ethics perspective by seeking to establish standards of conduct in a practical manner to promote better gamification research and development for educational settings. Normative ethics is a broad term that describes the moral reasoning (i.e., norms of conduct of what is acceptable or not) from multiple philosophical theories (e.g., consequentialism, deontological, and virtue ethics) and decision-making processes (e.g., planning, conducting and communicating) \cite{Farrimond2012}. 

Briefly describing some of these \textbf{philosophical theories}, the \textit{consequentialism} perspective follows a utilitarian rationale, prioritising societal over individual interests (i.e., favouring actions that benefit the majority of people), regardless of potential harms. In opposition, the \textit{deontological} perspective follows standards based on universal moral principles and duties to others (e.g., what would happen if everyone adhered to this standard?), disregarding how these codes vary according to their context and their unexpected results. At last, \textit{virtue ethics} perspective follows behaviours towards living a virtuous life by practising good traits (e.g., honesty, integrity) and emphasising the interdependency of human beings, which creates a moral obligation to care for dependent groups (e.g., children, older people) and to use emotional virtues (e.g., sensitivity, responsiveness) when interacting with people. Gamification scholars and practitioners usually focus on enhancing teaching-learning processes with motivational affordances to invoke psychological and behavioural outcomes, in which game elements are justified by their utility towards the so-called common good (i.e., consequentialism) \cite{Hamari2014}. At the same time, multiple studies discuss the so-called right way to design gamification, following a set of rules defined in a framework (i.e., deontology) \cite{Mora2017}. Relatively newer in the gamification field, the virtue ethics viewpoint invites scholars and practitioners to move from coercion to facilitating the best life and from instrumental perfection to critical transformation (i.e., ``a critical, transformative, socio-technical systems design practice for motivational affordances in the service of human flourishing'' -- eudaemonist virtue \cite{Deterding2014}).

In this sense, scholars and practitioners must ensure that gamification in educational settings focuses on students' flourishing, such as by providing a fulfilling and meaningful gameful learning experience, through a well-thought \textbf{decision-making process} throughout its research and development. For instance, in the \textit{planning phase}, those researching or developing gamification for educational domains must evaluate their competence in terms of skills and expertise (or collaborating with those with the necessary complementary abilities), as well as be familiar with relevant ethical guidelines for technology-assisted education through the lenses of cultural relativism and applicable legislation \cite{Aguinis2004}. As another example, the \textit{conduction phase} should follow principles of fairness, accountability, transparency and ethics (FATE) to avoid gamification designs that are exploitative or addictive \cite{Kim2015}, while ensuring that data is findable, accessible, interoperable and reusable (FAIR) \cite{Wilkinson2016}. As gamification scholars and practitioners, communicating the results of implementing such technology in education should be clear and comprehensive and respect students' privacy, whose data should be anonymised since the earlier stages of the gamified project \cite{Goswami2017}. Yet, a better understanding of the unethical issues in gamified education and a comprehensive set of guidelines to address them is still needed towards making ethical gamification development and making gamification research ethical. 

\section{Methodology}\label{sec:methology}

An umbrella review aims to summarise evidence from multiple research syntheses to provide an overall picture of findings for particular questions or phenomena \cite{Aromataris2015}. According to this goal, the umbrella review might also include studies regarding different conditions and populations \cite{Becker2011}. Furthermore, this methodology comprises a protocol detailing investigated research questions, inclusion criteria, search process, screening procedure and data extraction plan \cite{Becker2011}. Regarding this study, despite the existing multiple secondary studies on the intersection of gamification and ethics, there is still no comprehensive understanding of how to address and overcome unethical issues in the research and development of gamification in education. Towards this goal, this chapter describes an umbrella review that complies with the following:

\subsection{Research questions}\label{subsec:research-questions}

Towards understanding and finding ways to mitigate the ethical challenges of gamification when researching and developing such technologies, this chapter focuses on two main research questions:

\begin{enumerate}
    \item \textbf{How to make ethical gamification?} addressing how to design, implement and evaluate the effects of gamified educational applications towards living a virtuous life
    \item \textbf{How to make gamification ethical?} addressing how to plan, conduct and communicate the outcomes of gamification in educational settings 
\end{enumerate}

\subsection{Inclusion criteria}\label{subsec:inclusion-criteria}
Based on these research questions for a high-quality and assertive outcome, this umbrella review adopted the following inclusion criteria:

\begin{itemize}
    \item \textbf{Language:} Studies need to be written in English;
    \item \textbf{Venue:} Studies need to be published as Journal articles, Conference papers or Book chapters;
    \item \textbf{Methodology:} Studies need to conduct a secondary study;
    \item \textbf{Intervention:} Studies need to investigate gamification research, design or implementation; and
    \item \textbf{Outcome:} Studies need to tackle any ethical issues emerging from gamification. 
\end{itemize}

\subsection{Search process}\label{subsec:search-process}
The search string was set following the PICOC method, which defines the Population, Intervention, Comparison, Outcome and Context of the desired studies \cite{Wohlin2012}. In this chapter, the Population includes \textit{any secondary study}, using \textit{gamification} as the Intervention, and focusing on \textit{ethics} as the main Outcome. Based on the research questions, there is no Comparison to be made among the studies. No limitations were defined based on the Context of these studies, as investigating ethical aspects of gamification in a broader sense also contributes to a deeper understanding and anticipation of potential issues towards their mitigation in the educational domain, especially given that it also employs, reuses and benefits from overall research methodology and development. Therefore, the search was conducted on Scopus, which indexes many of the literature databases available, and considered studies that meet \textit{gamification AND ethic* AND (review OR systematic)} in their title, abstract, or keywords. The search was conducted in September 2022 and returned a total of 34 works. 

\subsection{Screening procedure}\label{subsec:screening-procedure}
Based on the selection criteria (described in Subsection \ref{subsec:inclusion-criteria}), the authors excluded studies based on their language (n = 1), venue (n = 7), methodology (n = 6), intervention (n = 7), and outcome (n = 8). Accordingly, a total of five secondary studies on the ethical issues and potential harms of gamification were included in this umbrella review \cite{Arora2021,Benner2021,Hassan2020,Humlung2019,Hyrynsalmi2017}.

\subsection{Data extraction plan}\label{subsec:data-extraction-plan}

Following the guidelines proposed by Aromataris et al. \cite{Aromataris2015}, this umbrella review extracted bibliometric information (i.e., citation details, objectives of the included reviews, review type, the context of the application, number of databases searched, publication range, number of studies included, and country of origin), and outcomes and implications reported that are related to the aforementioned research questions.

\section{Results}\label{sec:results}

In terms of bibliometric information, \textit{Arora and Razavian (2021) \cite{Arora2021}} conducted a systematic review to understand the ethical issues in the existing empirical work on the effects of gamification in health tracking. For this, the authors analysed 23 studies ranging between 2012 and 2021 from six different search engines (i.e., ACM Digital Library, IEEE Xplore, PhilPapers, PubMed, Scopus, and Web of Science). The second study, from \textit{Benner, Schöbel and Janson (2021) \cite{Benner2021}}, conducted a systematic review to understand the current ethical considerations in persuasive system design. The authors included 17 studies published between 2011 and 2020 from eight search engines (i.e., ACM Digital Library, AISel, Emerald, IEEE Xplore, JSTOR, PubMed, ScienceDirect and SpringerLink). Next, \textit{Hassan and Hamari (2020) \cite{Hassan2020}} conducted a systematic review to summarise what has been carried out on gamified e-participation. In this study, a total of 66 papers from Scopus were included, which the publication range is between 2012 and 2018. The fourth study, from \textit{Humlung and Haddara (2019) \cite{Humlung2019}}, was a systematic review of how to apply gamification in business as a means to create an innovative environment. While searching in Google Scholar, 19 studies published between 2013 and 2019 were analysed. The last included study, from \textit{Hyrynsalmi, Smed and Kimppa (2017) \cite{Hyrynsalmi2017}}, conducted a systematic review to understand the perceived negative side effects of applying gamification in a more general context. A total of 22 studies published between 2013 and 2016 were included in this study based on six search engines (i.e., ACM digital library, AISel, IEEE Xplore, ScienceDirect and Wiley Online Library). All these five systematic reviews were written by authors affiliated with European Universities (i.e., Finland \cite{Hassan2020,Hyrynsalmi2017}, The Netherlands \cite{Arora2021}, Germany and Switzerland \cite{Benner2021}, and Norway \cite{Humlung2019}). Details on their outcomes and implications are detailed below while addressing our research questions.

\subsection{How to make ethical gamification? (RQ1)}\label{subsec:development}

Towards achieving such critical and transformative gamification as aimed by virtue ethics, the current means to design, implement, and evaluate it must be rethought. In this sense, the ethical challenges found in the included studies were:

\textbf{\textit{Power dynamics and Paternalism.}} Shaping or reinforcing behaviours through persuasive technologies, such as gamification, \textit{without the proper communication} of the intentions behind gamification aligns with the consequentialism rationale, in which the benefits for the general audience are perceived to outweigh the potential harms \cite{Farrimond2012,Humlung2019}. More than a lack of communication, gamification has paternalistic characteristics that \textit{limit autonomy and freedom of choice} by positioning scholars and practitioners as authorities of the ``correct'' behaviour from the deontology rationale, while fostering the stigmatisation of those who are not able to meet these desired behaviours from gamification goals \cite{Arora2021,Benner2021}. Thus, despite being generally used to foster ``good'' habits, gamification design, implementation and evaluation usually focus on top-down approaches that patronise individuals instead of promoting autonomous and voluntary engagement \cite{Hassan2020}. While gamification scholars and practitioners are all susceptible to their own implicit biases,  consequentialism may take over again when the interest of a third party (e.g., companies, schools) contradicts and overcomes the original intention of helping people to achieve their own goals (e.g., dissonance, imbalance, conflict of interest), which now instead focuses on exploiting their real-world vulnerabilities \cite{Arora2021,Benner2021}. To address these challenges, \textit{any (digital) nudge should be disclosed to preserve people’ autonomy and freedom}, despite potential undesirable outcomes, given that those aware of this persuasion might react differently \cite{Benner2021}. On top of that, paternalistic characteristics can be avoided by \textit{educating people on nudging}, so that they would be aware of potential issues by themselves \cite{Hyrynsalmi2017}. Furthermore, gamification scholars and practitioners should \textit{include diverse stakeholders to account for multiple user voices during gamification design, implementation and evaluation processes} \cite{Hassan2020}, while ensuring that its persuasive effects are not misused to exploit people physically, financially, emotionally, or psychologically \cite{Benner2021}. In educational settings, gamification could also benefit from being based and aligned with transformative learning theories that allow a non-hierarchical dialogue among students and educators, so that individual needs are considered and learning becomes a more autonomous process where knowledge is promoted as collective construct \cite{Freire2000}.

\textbf{\textit{Lack of voluntarity.}} Following power dynamics from gamification being unfair to one party, another ethical issue relates to people \textit{feeling obliged to use gamified systems}. For instance, gamification might be deeply rooted in educational settings as an efficiency metric, while not being translated as beneficial for the students \cite{Arora2021,Benner2021}. Gamification not being entirely voluntary supports the power imbalance by intentionally or accidentally \textit{sugar-coating coercive practices} (i.e., by questionable means) and the \textit{reinforcement of desirable outcomes} (i.e., for questionable purposes) \cite{Hassan2020}. Examples of this can be seen, for instance, in a classroom where the educator gamifies a specific task that is more aligned with school needs (e.g., for external evaluation purposes) than individual learning meaningfulness, and the assessment of students is conditional on their participation, regardless if they have the means or desire to perform this task in a gamified way. To avoid such misconduct, \textit{gamification should include an opt-in design and proper anonymisation of people's information} \cite{Benner2021}. 

\textbf{\textit{Confidentiality issues.}} Another issue is the interaction between users and gamification providers. Mismanagement of the necessary communication between the parties can inflict fundamental ethical issues related to confidentiality. \textit{Anonymisation and providing information on what data and why they are collected while asking for explicit permission would allow information security and data privacy in gamified systems} \cite{Arora2021}. This would prevent dark patterns in the interaction design, such as cookies and consent default options favouring the gamification provider \cite{Benner2021}, intentionally luring people to share personal data through game elements, sharing or even selling personal data with third parties, and making people uncomfortable, anxious or any other psychological and emotional harm with their data being tracked or shared \cite{Arora2021}. In this sense, \textit{special attention should also be given when designing, implementing and evaluating gamification that might not ensure students' privacy} - such as avatars, challenges and competitions, which recognise and record information on students' characteristics, performance, and opponents \cite{Mavroeidi2020}.

\textbf{\textit{Cognitive manipulation}}. Gamification can also be a means to inhibit autonomy and undermine self-reflection in unjustifiable ways (e.g., distraction, addiction). For instance, gamification requiring instant reaction in some works and job positions, such as medics and firefighters, add unnecessary steps and \textbf{\textit{distractions}} that might cause dangers, remarkable losses, and threatening situations - overall physical and psychological harms \cite{Arora2021}. At the same time, the potential moral, ethical and legal challenges of gamblification require further investigation \cite{Benner2021}, while the overall \textbf{\textit{addiction}} might have detrimental effects on people, such as obsessively relying on game incentives or choosing goals that are potentially harmful (e.g., dieting based on what other people consider healthy) \cite{Arora2021}. Thus, \textit{gamification should provide safe restrictions or warnings against cognitive effects, allow autonomy and personalisation, and focus on facilitating internal motivation (e.g., self-determination and self-reward) for a sustainable behaviour change} \cite{Arora2021}. Furthermore, scholars and practitioners need to consider the context and target audience, such as adding gamification in products and services marketed for children or considering their impact on people with a history or tendency towards addiction \cite{Hyrynsalmi2017}. This, however, should not be a way to justify hyper-focus on specific target audiences, in which some people would have privilege over others, nor a means to reinforce stereotypes, in which different genders would have different colours or game elements in a reductionist way \cite{Arora2021}. Thus, \textit{gamification should be deeply aligned with intended learning outcomes to ensure it is not distracting students from the educational content}. At the same time, scholars and practitioners should \textit{understand the interaction of the students with the learning environment and its game elements to avoid reward dependency in the teaching-learning process} \cite{Andrade2016}.

\textbf{\textit{Social comparison}}. By drawing inspiration from games, gamification also allows competition and rivalry, potentially giving a sense of social overload and straining people \cite{Benner2021}. This might lead to a loss of motivation and a feeling of segregation for those who systematically perform worse than their counterparts and, furthermore, might lead to cheating. Thus, gamification scholars and practitioners \textit{should avoid giving people a sense of defeat while also transferring the responsibility to them by allowing cheating to some extent - with more autonomy in defining their own tasks, gamification supports a tolerant community of individual choice-making and acknowledges individual differences} \cite{Arora2021}. In educational settings, while considering that students have different learning styles, \textit{a tailor-made gamification} might be a good alternative to allow everyone to have fun and play the game even though the rules are not the same for everybody \cite{Klock2020}.

\subsection{How to make gamification ethical? (RQ2)}\label{subsec:research}

More than rethinking what aspects of gamified educational applications can be designed, implemented and evaluated in an unethical manner (RQ1), scholars and practitioners should follow ethical principles to promote ethical research and development of gamification (RQ2). Thus, gamification research and development are indivisible from ethics, as scholars' and practitioners' choices during the planning, conduction and communication have inherently ethical aspects \cite{Farrimond2012}. 

For instance, whenever \textbf{\textit{planning}} gamification research and development in educational settings, scholars and practitioners must consider which values and interests are promoted by the research questions and by the purpose of gamification in the educational application. Given supervisors' individual interests and business' own agendas, gamification investigation and execution might be loaded with contradictory assumptions (i.e., \textbf{\textit{conflict of interest}}) \cite{Farrimond2012}, which very much relate with \textit{power dynamics and paternalism} discussed in RQ1. Thus, scholars and practitioners must reflect on the gamification implications in educational settings before starting the project to ensure the research and development integrity \cite{Pimple2002}, especially regarding \textbf{\textit{sampling}} and \textbf{\textit{data collection methods}} through ethical lenses. Since the gamified education target audience tend to be quite broad and focuses on human beings, scholars and practitioners should clearly define whom the subjects are (i.e., who is included, but especially who is excluded and why), while avoiding presenting people in stereotyped ways (e.g., tailored gamification that generalises preferences based on a single characteristic). More than understanding ethical principles when involving people in research and development, scholars and practitioners should provide informed consent that ensures transparency and \textit{voluntarity}, clearly communicates potential risks and benefits of the participation, guarantees \textit{confidentiality}, privacy and anonymisation, prevents \textit{social comparison}, as well as affords special protections against \textit{addiction, distraction} and further needs for targeted populations (e.g., children, minority and elderly) \cite{Pimple2002}. 

Regarding \textbf{\textit{conducting}} gamification research and development in educational settings, gamification research and development should \textbf{\textit{ensure integrity and quality}} regardless of the chosen methodologies. For instance, involving participants from co-participatory approaches beyond the design phase allows a further co-production that promotes the inclusion of participants and their social worlds in the analysis (e.g., giving them a voice to agree or disagree with scholars and practitioners' understanding of their data, as opposing to \textit{power dynamics and paternalism}) and dissemination of the results (e.g., giving them credits for the co-production, while guaranteeing \textit{confidentiality}) \cite{Farrimond2012}. Overall, reliability and validity are some examples of measurements that ensure research quality in \textit{\textbf{quantitative methods}}, while transparency and data triangulation are examples of means to ensure quality when applying \textbf{\textit{qualitative methods}} \cite{Farrimond2012}. Nevertheless, doing ethical gamification research and development to the highest possible standard is not only a need but also a moral obligation. Scholars and practitioners should commit to following up-to-date ethic codes (e.g., APA Ethical Principles and Code of Conduct \footnote{\url{https://www.apa.org/ethics/code}}) to promote fairness \cite{Wilkinson2016} and avoid \textbf{\textit{misinterpretation}} or \textbf{\textit{misrepresentation}} of their gamification outcomes in educational settings.

Moreover, gamification outcomes need to be \textbf{\textit{communicated}} to relevant audiences, such as academia, industry and to the general public. Before that happens, it is essential to have an upfront discussion about the \textbf{\textit{authorship}} with scholars and practitioners involved in the research and development of gamification \cite{Farrimond2012}. While communicating, misconduct in all of its forms should be prevented: scholars and practitioners must not alter data (i.e., \textbf{\textit{falsification}}), nor publish data that were not actually collected (i.e., \textbf{\textit{fabrication}}), and mostly not steal other's ideas, methods or data without proper attribution (i.e., \textbf{\textit{plagiarism}}) \cite{Zigmond2002}. Finally, communication should be complete (i.e., avoid fragmentary publication) and comprehensive (i.e., with a sufficient description of methods, corrections, and retractions) \cite{Pimple2002}.

\section{Final remarks}\label{sec:conclusion}

This chapter addressed a series of ethical issues in gamification research and development, with a particular focus on the educational field. From the analysis of secondary studies, this umbrella review explored many ethical challenges of gamified educational applications and proposed potential solutions for future research and development. 

Towards designing, implementing and evaluating the effects of gamified educational applications towards living a virtuous life, we elaborated on \textbf{\textit{how to make ethical gamification}} (RQ1). Based on major unethical outcomes reported by the secondary studies, we propose that scholars and practitioners should ensure that digital nudging is properly disclosed to preserve people's autonomy and freedom, while educating people on nudging to raise awareness on potential issues from one's perspective. Gamification research and development should also involve diverse stakeholders to account for multiple user voices to further avoid \textit{power dynamics and paternalism}. As a mean to avoid coercive practices in face of the \textit{lack of voluntarity} from people in using gamification, gamification should include an opt-in design and proper anonymisation of user's information. Data anonymisation is also a good strategy to prevent \textit{confidentiality issues}, while providing information on what data and why they are collected when asking for explicit permission from the students. Gamification scholars and practitioners need also to be careful when designing and implementing some game elements (e.g., avatars, challenges and competitions) that might not ensure students' privacy. Gamification should provide safe restrictions or warnings against \textit{cognitive manipulation}, allowing autonomy and personalisation, and focusing on facilitating internal motivation. Towards preventing distractions and addiction, gamification needs to be aligned with the intended learning outcomes and understand students' interaction with the learning environment to avoid reward dependency. The autonomy is also important to stop \textit{social comparison}, by allowing students to define their own learning path or even promoting an autonomous tailoring system that understand students' individual preferences and needs. Towards planning, conducting and communicating the outcomes of gamification in educational settings, we elaborated on \textbf{\textit{how to make gamification ethical}} (RQ2), especially regarding conflict of interests, sampling and data collection methods, ensuring integrity and quality regardless of the chosen methodologies, misinterpretation and misrepresentation, authorship agreement, and misconduct through falsification, fabrication, and plagiarism.

However, this study has some limitations. As with any umbrella review, it was not possible to ensure the quality assessment of the included primary works, which may vary from one secondary study to another. As with any systematic study, our work and the secondary ones may have issues with the definition of the string, search engines and selection criteria, which may not have retrieved relevant papers during the search. Finally, since the screening of the study was conducted by one senior researcher in systematic reviews and gamification, but yet only one, we may have some false negative papers along the way. Because of that, our results could be slightly different from similar analysis approaches. Still, the ethical challenges in gamified educational research and development, as well as potential directions for future works, might be useful for scholars and practitioners as a first step towards promoting critical transformation of gamified educational applications and making them a tool to facilitate the best life.

\subsubsection*{Acknowledgements.} This work was supported by the Academy of Finland Flagship Programme [grant No 337653, Forest-Human-Machine Interplay (UNITE)] and the European Union’s Horizon 2020 research and innovation programme under the Marie Sklodowska-Curie [grant No 101029543, GamInclusive].

\end{document}